\documentclass[10pt,twocolumn]{article}
\usepackage{cite}
\usepackage{float}
\usepackage{pstricks}
\usepackage{color}
\usepackage[english]{babel}
\usepackage[spanish]{layout}
\usepackage[utf8]{inputenc}
\usepackage{layouts}
\usepackage{amsmath,amssymb}
\usepackage{enumerate}
\usepackage{textcomp}
\usepackage{url}
\usepackage{epstopdf}
\usepackage{graphicx} 
\usepackage{subfigure}
\usepackage{cancel}
\usepackage{hyperref}
\usepackage{cleveref}
 
\newcommand{\bc}{\begin{center}}
\newcommand{\ec}{\end{center}}

\newcommand{\be}{\begin{eqnarray}}
\newcommand{\ee}{\end{eqnarray}}
\newcommand{\bi}{\begin{itemize}}
\newcommand{\ei}{\end{itemize}}
 
\author{\small E. A. Reyes R.$^{1}$, A. R. Fazio$^{2}$\\
\footnotesize $^{1,2}$ \textsl{ Departamento de F\'{i}sica, Universidad Nacional de Colombia,}\\ \footnotesize \textsl{Ciudad Universitaria, Bogot\'{a} D.C., Colombia.}}
\date{\small \today}
\title{\textcolor{blue}{Comparison of the EFT Hybrid and Three-Loop Fixed-Order Calculations of the Lightest MSSM Higgs Boson Mass}}
\date{}
\begin{document}
\maketitle

\begin{abstract}
The lightest Higgs boson mass of the Minimal Supersymmetric Standard Model has been recently computed diagrammatically at the three-loop order in the whole supersymmetric parameters space of the SUSY-QCD sector. The code FeynHiggs combines one- and two-loop fixed-order with the effective-field-theory calculations for the same Higgs mass. The two numerical predictions agree considering the scenario of only one SUSY-scale and vanishing stop mixing parameter below 10 TeV. The agreement is improved by introducing an additional supersymmetric scale and a non-zero stop mixing. Additionally, the combined CMS/ATLAS Higgs mass value was used to derive an upper bound on the needed SUSY scale. In the considered scenario, values above the scale $12.5\pm1.2~\rm{TeV}$ are excluded. 
\end{abstract}   

\section{\large Introduction}
\label{intro}

The discovery by the ATLAS and CMS collaborations at the CERN Large Hadron Collider (LHC)~\cite{ATLAS, CMS} of a bosonic particle, with properties which are compatible with those predicted for the Higgs boson of the Standard Model (SM), represents a significant progress in our understanding of the electroweak symmetry breaking mechanism. The SM is theoretically consistent with the inclusion of a $125~\rm{GeV}$ Higgs boson, in the sense that no Landau pole emerges, also if the model is extrapolated up to the Planck scale ($\Lambda_P\approx 10^{18}~\rm{GeV}$), where one has to accept a meta-stable vacuum and an unnatural high amount of fine-tuning ($10^{34}$) for the prediction of the Higgs boson mass at the electroweak scale ($\Lambda_{EW}\approx 10^2~\rm{GeV}$)~\cite{Isidori, Buttazzo, Bezrukov, Kniehl, kniehl2}. However, there are still several puzzles that remain unsolved by the SM dynamics. The hierarchy problem, the neutrino oscillation, the identification of the dark matter, the baryon asymmetry, among others, are all left unanswered and require new physics beyond the Standard Model. The minimal super-symmetric extension of the SM (MSSM) is the best motivated and the most intensively studied framework of new physics, providing a widely amount of precise predictions for experimental phenomena at the TeV scale~\cite{Nilles,HabernKane}. In most scenarios that are phenomenologically relevant~\cite{Stal, Carena, Bagnaschietal,HST} the LHC measured value, $M_h^{exp}=125.09\pm 0.24~\rm{GeV}$~\cite{MhMass,MhCMS,MhATLAS}, is associated with the lightest CP-even Higgs boson mass ($M_h$) which is theoretically predicted with great accuracy in the MSSM. Up to now, the dominant quantum corrections to $M_h$ have been computed at one-loop~\cite{Chankowski,Dabelstein,Pierce,Frank}, two-loop~\cite{alphaalphas,alphabalphas,alphatalphas,Carena2,CarenaHollik,MartinHiggs,Sophia,
DegrassiMSSM,SophiaPaser} and three-loop \cite{Harlander1,Harlander2,Kant,Harlander3,Edilson} level using the Feynman diagrammatic (FD) and the effective potential (EP) approaches. These MSSM predictions can accommodate the measured Higgs mass value of $125~\rm{GeV}$ and are consistent with the similarities of the measured Higgs couplings to those in the SM~\cite{SMCouplings}. Effective field theory (EFT) methods have been also considered to resum large-logarithms in case of a large mass hierarchy between $\Lambda_{EW}$ and the SUSY scale ($M_{SUSY}$)~\cite{Draper,Lee,Villadoro,Bagnaschi}. In particular, for values of $M_{SUSY}$ above a critical point where the fixed-order and EFT combined uncertainties are equal, the EFT computation is more accurate and therefore the usage of the SM~\cite{Allanach} or a two-Higgs-doublet-model (THDM)~\cite{Bahl2} as effective theories below the SUSY scale is preferred. Both the fixed-order and the EFT results are implemented in several publicly available codes. For the diagrammatic fixed-order calculations there are the programs SoftSUSY~\cite{Allanach2}, SUSPECT~\cite{SUSPECT}, CPSuperH~\cite{CPSuperH} and H3m~\cite{Harlander3}. Pure EFT calculations are implemented in SUSYHD~\cite{Villadoro} and MhEFT~\cite{MhEFT}. Moreover, different hybrid methods that combine both approaches have been recently developed in order to take profit of the features of each one. FlexibleSUSY~\cite{FlexibleSUSY}, based on SARAH~\cite{SARAH}, implements a hybrid method called Flexible-EFT-Higgs~\cite{Athron}. This approach was also included into the program SPheno~\cite{SPheno1,SPheno2}. \\ A hybrid method different from the one pursued in Flexible-EFT-Higgs has been implemented in FeynHiggs~\cite{FeynHiggs1,FeynHiggs2}. There are also in literature detailed numerical comparisons between the different diagrammatic, EFT and hybrid codes. In~\cite{Athron} it is discussed in details how the hybrid method Flexible-EFT-Higgs compares to the other EFT and diagrammatic codes. Finally, several numerical comparisons of the hybrid approach implemented in FeynHiggs to the pure EFT calculations have been studied in~\cite{Villadoro,Athron,SPheno2}. Those papers reported surprising non-negligible numerical differences between FeynHiggs and pure EFT codes for the prediction of $M_h$ at large SUSY scales. The observed differences come mainly from three sources. The scheme conversion of input parameters from OS to $\overline{DR}$, which can lead to large shifts due to uncontrolled higher-order terms. Unwanted effects from incomplete cancellations with subloop renormalization contributions in the determinations of the Higgs propagator pole and different
parametrizations of non-logarithmic terms. After performing the corresponding corrections, FeynHiggs results are in very good agreement with the results of SUSYHD~\cite{Bahl3}. \\ For the present study we  decided to use the fixed-order and EFT hybrid calculations currently included in FeynHiggs, which seems to be in a very good agreement with the other fixed-order and EFT codes and gives a reliable three-loop predictions of the Higgs boson mass for large SUSY scales, in order to provide a numerical comparison of our three-loop fixed-order predictions of the ligthest MSSM Higgs boson mass reported in~\cite{Edilson} with the fixed-order and EFT hybrid results found in literature. As the effects of the large logarithms are expected to become relevant when $M_{SUSY}$ grows, it is natural to ask how large $M_{SUSY}$ can be. We therefore provide in this article a phenomenological analysis about the compatibility of the experimental observations at the LHC for the Higgs boson mass and the region of parameters in the specific MSSM considered scenario to find an upper bound on the needed $M_{SUSY}$.                        

\section{ \large Three-Loop Fixed-Order Calculation of~${M_h}$}
\label{sec-1}

\begin{figure}[h]
\centering
\includegraphics[width=18pc]{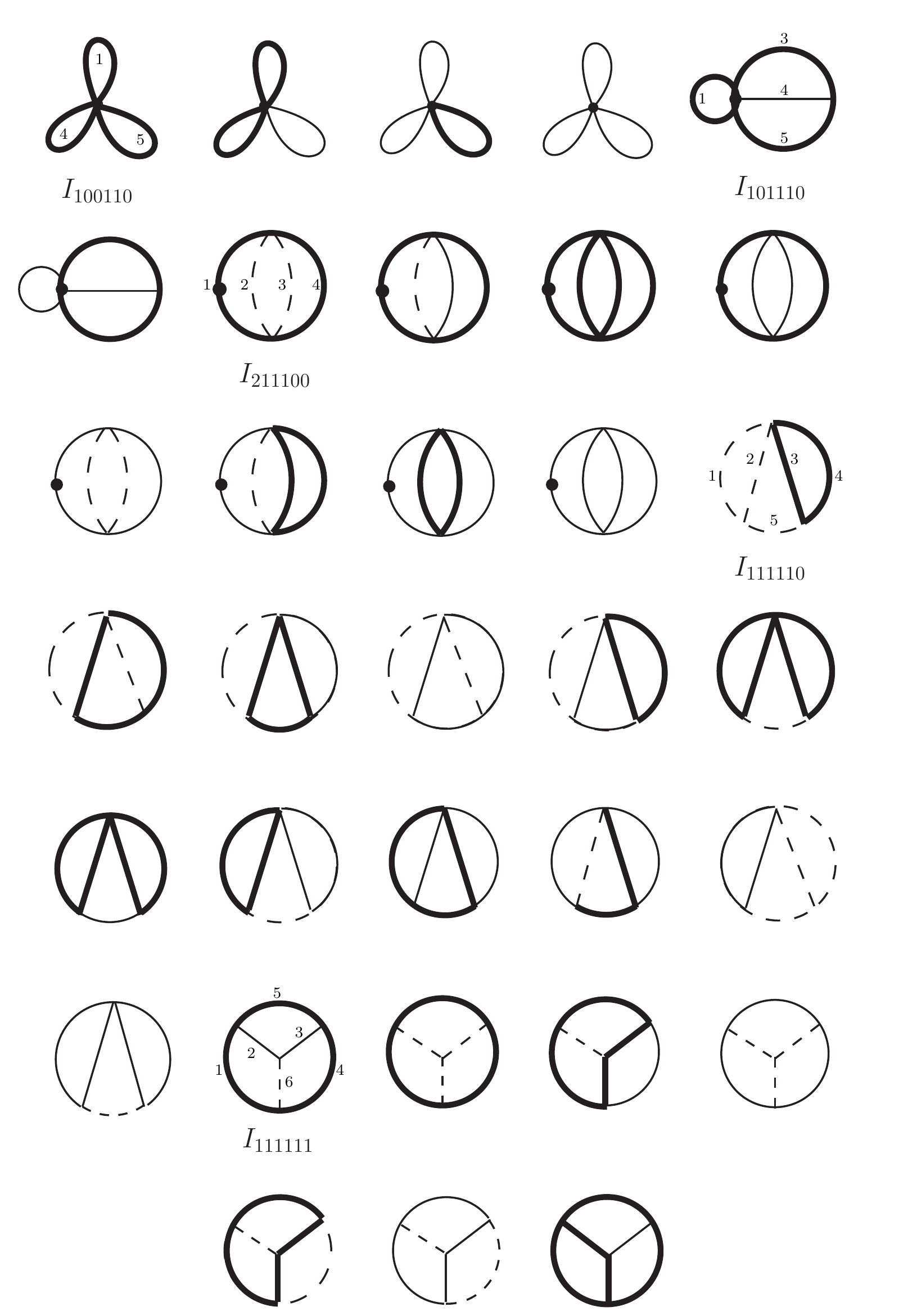}
\caption{\small{Basis of three-loop Master Integrals. The dashed line represents a massless propagator. The thin solid line is the propagator with a mass at the electroweak scale $M_t$ and the thick solid line depicts the propagator involving the SUSY scale $M_{SUSY}$.}}
\label{MasterIMS}
\end{figure}

In contrast to the SM, the Higgs sector of the MSSM with real parameters (rMSSM) contains two complex doublets with opposite hyper-charges 
\begin{eqnarray}
\tiny
H_{1}=\left(\begin{array}{c}
H_{1}^{0}+\frac{v_{1}}{\sqrt{2}}\\
H_{1}^{-}
\end{array}\right) & \tiny \rm{and} & \tiny H_{2}=\left(\begin{array}{c}
H_{2}^{+}\\
H_{2}^{0}+\frac{v_{2}}{\sqrt{2}}
\end{array}\right), \label{eq:Higgsdoublets}
\end{eqnarray}
where the neutral components, $H_{1,2}^0$ fluctuate around the vacuum expectation values (vevs) $v_{1,2}$. In the physical basis there are five Higgs bosons, three of them are neutral: the lightest ($h$) and heavy ($H$) CP-even Higgs bosons and the CP-odd Higgs boson ($A$). The other two, $H^{\pm}$, are charged and vev-less. Besides the SM electroweak boson masses, the rMSSM Higgs sector is parametrized in terms of two additional parameters: the mass of the CP-odd Higgs boson ($M_{A}$) and $tan\beta$, which is the ratio of the two vevs, $v_1/v_2$. The masses of the CP-even Higgs boson particles, $h$ and $H$, follow as predictions. \\ We focus in this section on the prediction of the lightest Higgs boson mass, $M_h$, at three-loop accuracy using a fixed-order FD computation which is based on the calculation of Higgs self-energy corrections at the given perturbative order. In this approach, the renormalized CP-even Higgs boson masses are obtained by finding the zeros of the determinant of the inverse propagator matrix (poles equation)
\begin{eqnarray}
\tiny
\left(\Delta_{H}\right)^{-1}=-i\left(\begin{array}{cc}
p^{2}-m_{H}^2 + {\displaystyle\sum_{l=1}^{3}} \widehat{\prod}_{HH}^{(l)} & {\displaystyle\sum_{l=1}^{3}}\widehat{\prod}_{hH}^{(l)}\\
{\displaystyle\sum_{l=1}^{3}}\widehat{\prod}_{hH}^{(l)} & p^{2}-m_{h}^2 + {\displaystyle\sum_{l=1}^{3}}\widehat{\prod}_{hh}^{(l)}
\end{array}\right), \label{eq:inversematrix}
\end{eqnarray} 
where $m_{h}$ and $m_{H}$ denote the tree-level mass of $h$ and $H$ respectively and 
\begin{eqnarray}
\tiny
\left.\widehat{\prod}\right._{ij}^{(l)}=\left.\prod\right._{ij}^{(l)}~-~\delta^{(l)}M_{ij}^{2} & \tiny ; & \tiny i,\: j = h,\: H, \label{eq:SEandRC}
\end{eqnarray}
are the corresponding $l$-loop renormalized self-energies. A particular feature of the rMSSM is the large size of the higher order quantum corrections to masses and couplings. They can lead to a considerably large shift on the value of the Higgs boson mass, where the bulk of the corrections comes from the SUSY-QCD sector of the Lagrangian. Thus, the dominant contributions to $\left.\widehat{\prod}\right._{ij}$ in eq. (\ref{eq:SEandRC}) involve the SM parameters $h_t$ (top Yukawa coupling), $M_t$ (top quark mass), $\alpha_s$ (strong coupling constant) and the MSSM parameters $M_{\tilde{g}}$ (gluino mass), $\theta_t$ (stop mixing angle), $\tilde{m}_{q_{1,2}}$ (squark masses) and $A_q$ (soft breaking parameters) where $q=u,d,t,b,c,s$. \\ Concerning the renormalization of the self-energy corrections, that is to say, the determination of the mass counter-terms $\delta^{(l)}M_{ij}^{2}$, we follow the mixed OS/$\overline{\rm{DR}}$ scheme defined in~\cite{Edilson}. Thus, the electroweak gaugeless limit at O($\alpha_t\alpha_s^2$) and the approximation of zero external momentum are assumed. As a consequence, we have avoided dealing with the Higgs wave function renormalization and also with the renormalization of $tan\beta$. Moreover, $v_{1,2}$ are defined as the minima of the full effective potential and therefore the tadpoles are renormalized on-shell according to the conditions:
\begin{eqnarray}
\tiny
T_{1,2}^{tree}=0, & \tiny \delta^{(l)}T_{1,2}=-T_{1,2}^{(l)},\label{eq:T(OS)}
\end{eqnarray}
where $T_{j}^{(l)}$ is the $l$-loop Higgs tadpole contribution. We have also imposed an on-shell renormalization to the $A$-boson mass, 
\begin{eqnarray}
\tiny
\delta^{(l)}M_{AA}^2=Re\left[\left.\prod\right._{AA}^{(l)}\left(M_{A}^{2}\right)\right]. \label{eq:MA(OS)}
\end{eqnarray}
The three-loop corrections of O($\alpha_t\alpha_s^2$) also include the O($\alpha_s$) contributions to the one-loop counter-terms coming from the renormalization of the gluino mass, the top quark mass, the squark masses and the stop mixing angles in the $\overline{DR}$ scheme, as well as the two-loop $\overline{DR}$ renormalization of the top mass, the stop masses and stop mixing angles at O($\alpha_s^2$).\\ For the purposes of this article we have chosen a degenerate single-scale scenario  where all the super-symmetric masses are set equal to an effective scale $M_{SUSY}$,
\begin{eqnarray}
\tiny
M_{L,R} = M_{\tilde{g}} = M_A =\mu = M_{SUSY}. \label{eq:scales}
\end{eqnarray}
Here $\mu$ is the Higgsino mass and $M_{L,R}$ are the soft SUSY-breaking masses. We have also identified the lightest Higgs boson $h$ as the SM-like Higgs boson and therefore we have assumed the decoupling limit, $M_A = M_{SUSY} \gg M_t$. This degenerate scenario in the decoupling limit is known as the ``heavy SUSY" limit. As a consequence, the three-loop self-energy corrections to $m_{h,H}^{2}$ can approximately be obtained as a superposition of the $33$ vacuum integrals depicted in Fig. \ref{MasterIMS} with coefficients that are functions of the kinematic invariants and the space-time dimension.       
Each diagram of the basis in Fig. \ref{MasterIMS} represents a three-loop Master Integral of the form
\begin{eqnarray}
\tiny
I_{v_{1}\ldots v_{6}}=i\frac{e^{3\gamma_{E}\varepsilon}}{\pi^{3D/2}}\int{\displaystyle \prod_{l=1}^{3}}d^{D}q_{l}\left[{\displaystyle \prod_{j=1}^{6}}\frac{1}{P_{j}^{n_{j}}}\right], \label{eq:MI}
\end{eqnarray}
where
\begin{eqnarray*}
\tiny P_{1}=q_{1}^{2}-m_{1}^{2},~~~~~~~ & \tiny P_{2}=\left(q_{1}-q_{2}\right)^{2}-m_{2}^{2}, \\
\tiny P_{3}=\left(q_{2}-q_{3}\right)^{2}-m_{3}^{2}, & \tiny P_{4}=q_{3}^{2}-m_{4}^{2}, \\ 
\tiny P_{5}=q_{2}^{2}-m_{5}^{2},~~~~~~~ & \tiny P_{6}=\left(q_{1}-q_{3}\right)^{2}-m_{6}^{2}.
\end{eqnarray*}
There are two scales involved, the electroweak scale $M_t$, whose associated propagator is represented with a thin solid line and the super-symmetric scale $M_{SUSY}$ represented with a thick solid line. Massless propagators are represented with a dashed line. This basis was obtained using the integration by parts (IBP) method implemented in the code Reduze~\cite{Reduze}. Main part of the diagrams shown on Fig.~\ref{MasterIMS} have been analytically evaluated in \cite{MIs1,MIs2,MIs3,MIs4,MIs5,MIs6,MIs7,MIs8}. The numerical evaluation of the basis integrals was done with the programs TVID~\cite{Freitas1, Freitas2} and SecDec~\cite{SecDec}. In particular, the integral $I_{211100}$ requires a Laurent expansion up to first order in $\varepsilon$. The evanescent terms of O($\varepsilon^1$) was numerically evaluated with the help of SecDec. 

\section{\large EFT Hybrid Calculation of~$\rm{M_h}$}
\label{sec-2}
When there is a large mass hierarchy between the electroweak scale and the scale of the SUSY particles, the fixed-order computations of the Higgs self-energy corrections contain large logarithms that can spoil the convergence of the perturbative expansion and yield unreliable predictions of the Higgs boson masses. A fixed-order computation is thus recommended for low values of $M_{SUSY}$ not separated too much from $M_t$. There is an alternative approach to calculate $M_h$ which yield accurate results for high SUSY scales. This approach is based on the EFT techniques~\cite{Villadoro,Heidi} and allows the resummation of the large logarithmic terms and the incorporation of higher-order contributions beyond the order of the fixed-order diagrammatic calculations. In the heavy SUSY limit the low-scale EFT below $M_{SUSY}$ is the SM. It requires just one EFT coupling, the effective Higgs self coupling $\lambda$, which correlates the high scale $M_{SUSY}$ and the low scale $M_t$ through the renormalization group equations (RGEs) and captures radiative corrections of the form 
\begin{eqnarray}
\tiny
\alpha_j^{n+m-1}log^{n}\left(M_{SUSY}/M_t\right) & \tiny ; & \tiny j = \lambda,h_t,g_s,..., \label{eq:largelog}
\end{eqnarray}
for any $n$, by using the $m$-loop beta functions of $\alpha_j$, into the running coupling $\lambda(Q)$. In order to get a SM running Higgs mass in the $\overline{MS}$ scheme at the scale $M_t$, one has to multiply $\lambda(M_t)$ by $2v^2(M_t)$, where $v(M_t)\approx 246~\rm{GeV}$ is the $\overline{MS}$ vev evaluated at $M_t$. The physical Higgs mass requires to solve the pole equation   
\begin{eqnarray}
\tiny
p^2 - 2\lambda(M_t)v^2(M_t) + \widetilde{\prod}_{hh}^{SM}\left(p^2\right) = 0, \label{eq:SMpoleEq}
\end{eqnarray}
with the SM Higgs boson self-energy,
\begin{eqnarray}
\tiny
\widetilde{\prod}_{hh}^{SM}\left(p^2\right) = \left[\left.\prod\right._{hh}^{SM}\left(p^2\right)-\frac{1}{\sqrt{2}v}T_h^{SM}\right]_{fin}, \label{eq:SMSE}
\end{eqnarray}
renormalized in the $\overline{MS}$ scheme but with the Higgs tadpoles renormalized to zero, i.e. $\delta T_h^{SM} = -T_h^{SM}$. \\ As higher dimensional operators are not included into the effective Lagrangian, the contributions suppressed by the heavy scale $M_{SUSY}$ are not considered. Consequently, the EFT calculation is less accurate than the fixed-order one for low SUSY scales. The fixed-order calculation is more accurate below a critical SUSY mass scale, estimated to be about $M_{SUSY}^{C}\approx 1.2~\rm{TeV}$ in~\cite{Allanach}, whereas above that scale the EFT calculation is more accurate. \\ In the latest released version of FeynHiggs~\cite{FeynHiggs3} both approaches are combined in order to supplement the full one-loop, leading and sub-leading two-loop diagrammatic results with a resummation of the leading + next to leading (LL+NLL)~\cite{THahn} and next to next to leading (NNLL)~\cite{Bahl1} logarithmic contributions coming from the top/stop sector. 
For the resummation of large logarithms up to NLL two-loop RGEs and one-loop
matching conditions are needed, accordingly, the resummation up to NNLL requires three-loop RGEs and two-loop matching conditions. The hybrid results obtained from the combination of the two approaches are added into the pole equation of the full MSSM
\begin{eqnarray}
\tiny
p^2 - m_h^2 + \widetilde{\prod}_{hh}\left(p^2\right) + \Delta_{hh}^{log} = 0, \label{eq:MSSMpoleEq}
\end{eqnarray}     
through the shift $\Delta_{hh}^{log}$ which contains the resummed large logarithms from the EFT as well as the logarithmic terms already present in the fixed-order Higgs self-energies,
\begin{eqnarray}
\tiny
\Delta_{hh}^{log} = - \left[2\lambda(M_t)v^2(M_t)\right]_{log} - \left[ \widetilde{\prod}_{hh}\left(m_h^2\right) \right]_{log}. \label{eq:hhShift}
\end{eqnarray} 
The subscript $"log"$ means that only logarithmic terms are considered. The logarithms in the Higgs self-energy appear explicitly only after expanding in $v/M_{SUSY}$. This subtraction term ensures that the one- and two-loop logarithms, already contained in the fixed-order FD computation, are not counted twice. In general the higher-order logarithms obtained from the EFT and the hybrid approaches are not the same because the determination of the poles of the propagators (eq. \ref{eq:SMpoleEq} and eq. \ref{eq:MSSMpoleEq}) are performed in different models. However, this difference, which comes from the momentum dependence of the two-loop order non-SM contributions to the Higgs self-energy, cancels out with contributions coming from the subloop renormalization in the heavy SUSY limit, as was explicitly shown in~\cite{Bahl3}. Besides the unwanted effects from incomplete cancellations in the determination of the Higgs propagator pole, the effects due to non-logarithmic terms and its parametrization as well as the higher-order terms coming from the scheme conversion between $OS$ and $\overline{DR}$ parameters are all included into FeynHiggs~2.14~\cite{FeynHiggs3}.         
 
\section{\large Numerical Results}
\label{sec-3}

In this section we present a numerical comparison of our three-loop fixed-order predictions of $M_h$ to the numerical predictions coming from the new version of FeynHiggs. 
We have chosen a $\overline{DR}$ renormalization of the stop sector with the renormalization scale set to $\mu_r=M_{SUSY}$, which is equivalent to set $Q_t = -1$ in FeynHiggs. The one-/two-loop fixed-order and the EFT-hybrid FeynHiggs predictions are fixed such that the full MSSM is considered (mssmpart=4) in its real version (higgsmix=2, tlCplxApprox=0), no approximation is taken for the one-loop result (p2approx=4) and the $O(tan^n \beta)$ corrections are resummed (botResum=1). In particular, when the resummation of the large logarithms is included, we use the full LL, NLL and NNLL resummation (looplevel=2, loglevel=3). The top quark mass is renormalized in the SM $\overline{MS}$ scheme at NNLO (runningMT=1) since for loglevel different from zero a $\overline{DR}$ renormalization is not allowed. The input flags of FeynHiggs 2.14.3 are explicitly indicated, for more details the online manual of the code can be consulted at~\cite{manual}. To obtain the pole mass $M_h$ at three-loop level in the fixed-order approach, we have introduced the $O(\alpha_t\alpha_s^2)$ corrections as constant shifts in the FeynHiggs 1-loop + 2-loop Higgs renormalized self-energies~(looplevel=2 and loglevel=0) with the help of the function FHAddSelf but in this case we have used a $\overline{DR}$ renormalization of the top quark mass (runningMT=3).
\begin{figure}[H]
\centering
\includegraphics[width=18pc]{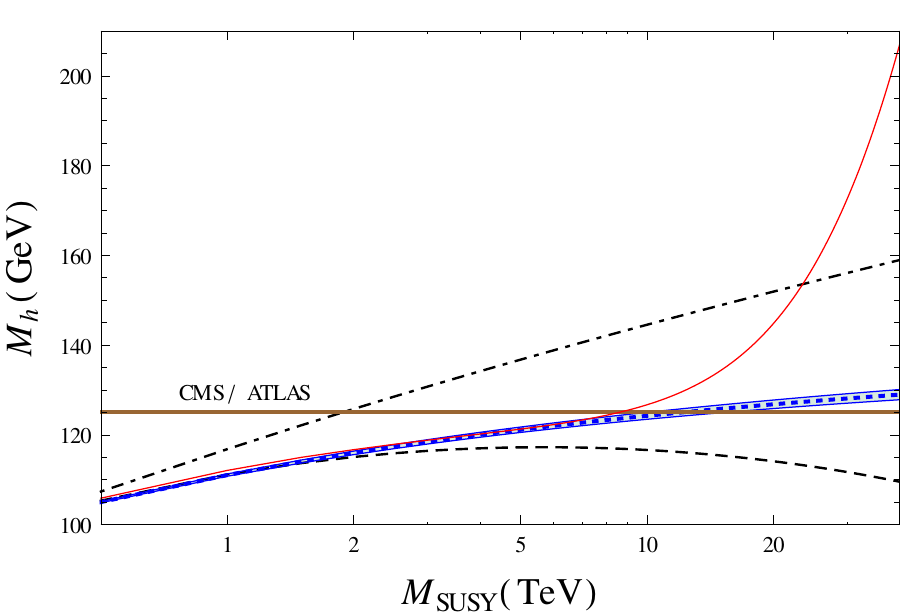}  \\
\includegraphics[width=17.9pc]{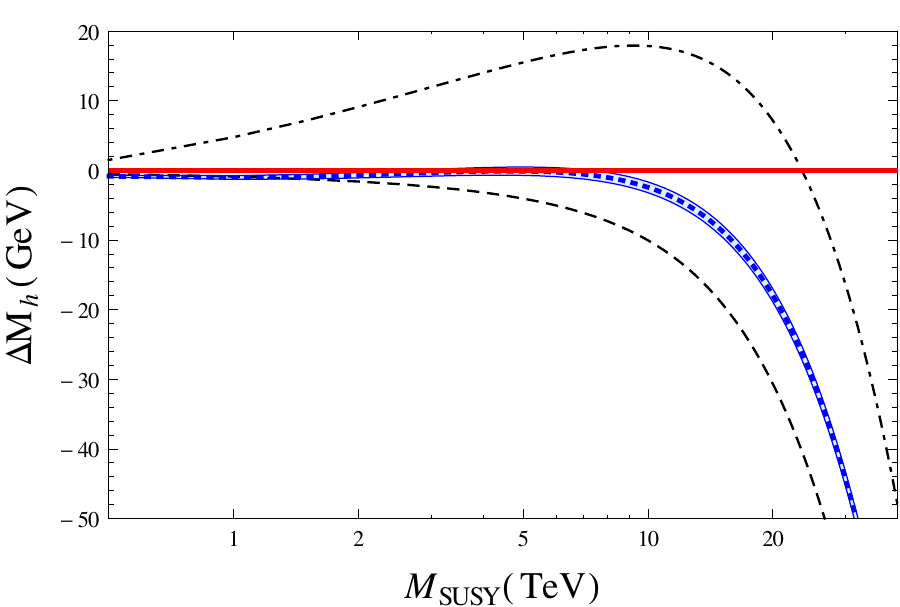} 
\caption{\small{Comparison of the $M_h$ predictions of FeynHiggs with the three-loop fixed-order computation of $M_h$ at $O(\alpha_t\alpha_s^2)$ in the heavy SUSY limit. The dot-dashed and the dashed lines are the fixed-order results of FeynHiggs at one and two -loop level respectively. The blue dotted line contains the NNLL resummation of the large logarithms in FeynHiggs. The blue band corresponds to the uncertainty in the NNLL prediction taken from FeynHiggs. The brown band is the CMS/ATLAS Higgs boson mass, $M_h^{exp}=125.09 \pm 0.24~\rm{GeV}$. The red solid line represents our three-loop fixed-order predictions. Up: Dependence of $M_h$ on the super-symmetric scale $M_{SUSY}$ for a vanishing stop mixing, $X_t/M_{SUSY}=0$. Down: Numerical differences between the FeynHiggs predictions and the three-loop fixed-order predictions of~$M_h$.}}
\label{Fig1}
\end{figure}
We start by considering the FeynHiggs fixed-order, FeynHiggs NNLL hybrid and three-loop $O(\alpha_t\alpha_s^2)$ predictions. The upper plot of Fig. \ref{Fig1} shows the dependence of $M_h$ on $M_{SUSY}$ for a vanishing stop mixing, $X_t/M_{SUSY}=0$, at the kinematic point $A_{e,\mu,\tau,u,d,c,s,b}=0$ and $tan\beta=10$, whereas the lower plot shows the numerical differences between all the considered FeynHiggs results and the $O(\alpha_t\alpha_s^2)$ prediction of $M_h$. In order to draw these plots we have adopted the heavy SUSY limit (eq. \ref{eq:scales}) and we have followed the next conventions. The one and two-loop fixed-order results of FeynHiggs are represented with the dot-dashed and the dashed lines respectively. The blue dotted line contains, in addition, the resummation of the large logarithms up to NNLL order. The blue band corresponds to the uncertainty in the NNLL prediction computed with the help of the FeynHiggs function FHUncertainties for the flag choise: mssmpart = 4, looplevel = 2, loglevel = 3, runningMT = 1.  In principle three effects are taken into account: i) the variation of the renormalization scale from $M_t/2$ to $2M_t$, ii) the use of $M_t^{pole}$ instead of $M_t^{run}$ in the two-loop corrections and iii) the exclusion of higher order resummation effects in $M_b$. The brown band is the experimental Higgs boson mass and its corresponding uncertainty, we have adopted the combined CMS/ATLAS result of the RUN 1 at the LHC, $M_h^{exp}=125.09 \pm 0.24~\rm{GeV}$~\cite{MhMass}, since there is not yet an official combined result for RUN~2~\cite{MhCMS,MhATLAS} observations. Finally, the red solid line contains our three-loop fixed-order corrections. \\ 
The first thing to note here (and also in Fig. \ref{Fig2}) is that the higher-order large logarithms coming from the EFT hybrid approach at NNLL level produce a growing positive shift on the two-loop predictions reaching a size of about 20 GeV for $M_{SUSY}=40~\rm{TeV}$. Additionally, the NNLL predictions are in a very good agreement with the three-loop $O(\alpha_t\alpha_s^2)$ results for $M_{SUSY}$ less than the value $M_{SUSY} \lesssim 10~{\rm TeV}$. 
On the lower graph of Fig.~\ref{Fig1} one can see that in the region $2.2~{\rm TeV} \lesssim M_{SUSY} \lesssim 7.4~{\rm TeV}$ there is an approximately constant difference of about $0.2~\rm{GeV}$ between the red solid and the blue dotted line which is within the theoretical uncertainty (blue band) estimated to be about 0.6~GeV. Below this region the agreement is still good with a numerical difference of at most $1~\rm{GeV}$. However, for scales above 10~TeV the effects of the large logarithms in the red curve start to be relevant, the difference between the two results rapidly increases up to $\sim 21~\rm{GeV}$ when $M_{SUSY}$ grows to up to $20~\rm{TeV}$ and grows monotonically reaching 78 GeV at $M_{SUSY}=40~\rm{TeV}$. This pronounced behaviour depends crucially on our election of the input parameters $\mu_r,~M_{\tilde{g}}~{\rm and}~X_t$. The presence of $n$-loop logarithms of the form $log^{n}\left(M_{SUSY}/M_t \right)$ in the master integrals of Fig.~\ref{MasterIMS} can introduce additional large contributions in the three-loop predictions of $M_h$. \\  
In Fig.~\ref{Fig2} the heavy SUSY limit has been smoothed to include an additional SUSY scale, the gluino mass $M_{\tilde{g}}$, in the fixed-order results. The NNLL resummation procedure is restricted to the case of $M_{\tilde{g}}$ equal to $M_{SUSY}$ since three-loop RGEs for an appropriate extension of the Standard Model with the gluino as additional fermion \cite { Bahl1}, for instance as a singlet of the gauge group, are not included in FeynHiggs. However, the main contributions sensitive to the gluino mass are captured by the two-loop result, the numerical effects due to a gluino threshold is numerically small and can be safely neglected, as was shown in~\cite{Bahl1}. We have considered a gluino mass of $M_{\tilde{g}}=1.5~\rm{TeV}$. The inclusion of this additional scale produces sizeable differences between the $O(\alpha_t\alpha_s^2)$ and the  NNLL results. For small SUSY scales below $\sim 3.5~\rm{TeV}$ the difference is always less than $1.3~\rm{GeV}$. For large SUSY scales ($M_{SUSY}>3.5~\rm{TeV}$) this difference grows to a maximum value of 4 GeV when $M_{SUSY}=20~\rm{TeV}$. Nevertheless, the numerical effect of the large logarithms in the red curve is reduced by a factor of around 5 regarding the results shown in Fig. \ref{Fig1}. Finally, we have studied the dependence of the NNLL and three-loop $M_h$ predictions on the stop mixing parameter $X_t$ in the heavy SUSY limit.
\begin{figure}
\centering
\includegraphics[width=18pc]{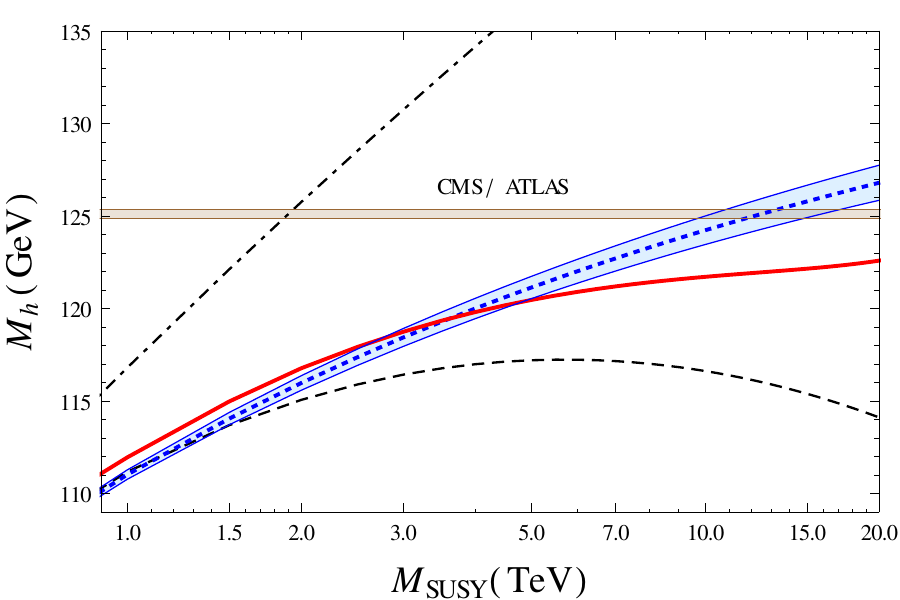}  \\
\includegraphics[width=18pc]{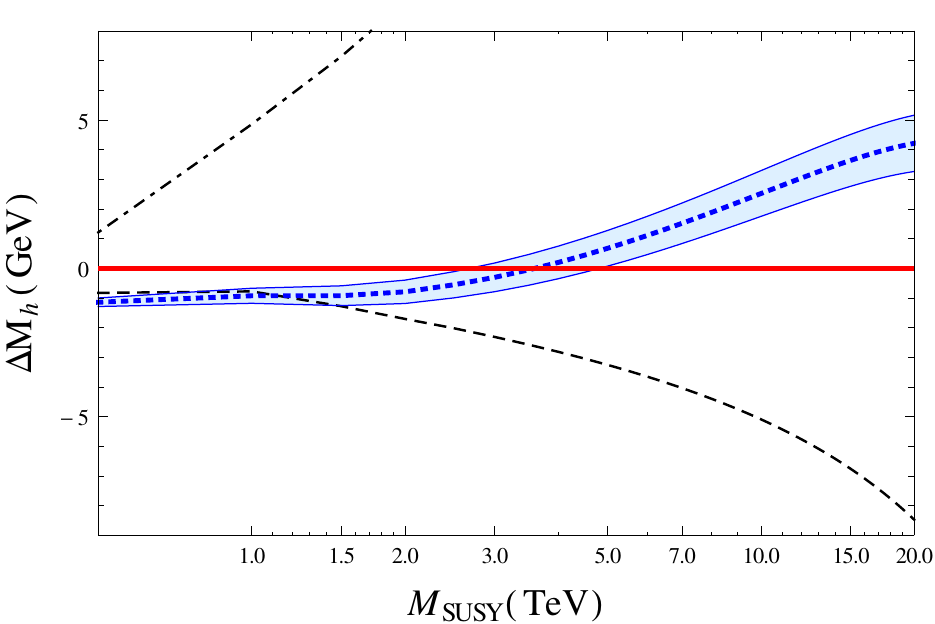}
\caption{\small{Numerical comparison of the $M_h$ predictions in a scenario where $M_{\tilde{g}}=1.5~\rm{TeV}$ and $X_t/M_{SUSY}=0$. These plots follow the same conventions as in the Figure \ref{Fig1}. Up: Evolution of $M_h$ as a function of $M_{SUSY}$. Down: Differences between the three-loop fixed-order and the FeynHiggs predictions.}}
\label{Fig2}
\end{figure}
In Fig. \ref{Fig3} we increased the value of $X_t/M_{SUSY}$ from 0.2 (thin curves) to 2.4 (thick curves). We observe that for high energy scales above $M_{SUSY} \gtrsim 10~\rm{TeV}$ the agreement between the two predictions and therefore the effect of the large logarithms on the red curves improves when $X_t/M_{SUSY}$ increases up to $1.5$, where the numerical size of the large logarithmic contributions in the $O(\alpha_t\alpha_s^2)$ results is reduced by a factor of about 7 regarding the non-mixing scenario ($X_t=0$). At the kinematic point $X_t/M_{SUSY}=1.5$, the curvature of $M_h$ as a function of $X_t$ changes its sign and therefore $\Delta M_h$ starts to increase again for even higher values ($X_t/M_{SUSY}>1.5$) reaching the maximum difference in the critical mixing $X_t/M_{SUSY}=2.4$ (thickest lines in Fig. \ref{Fig3}) which is another inflection point of $M_h(X_t)$ where the prediction of $M_h$ takes its higher value.
\begin{figure}
\centering
\includegraphics[width=18pc]{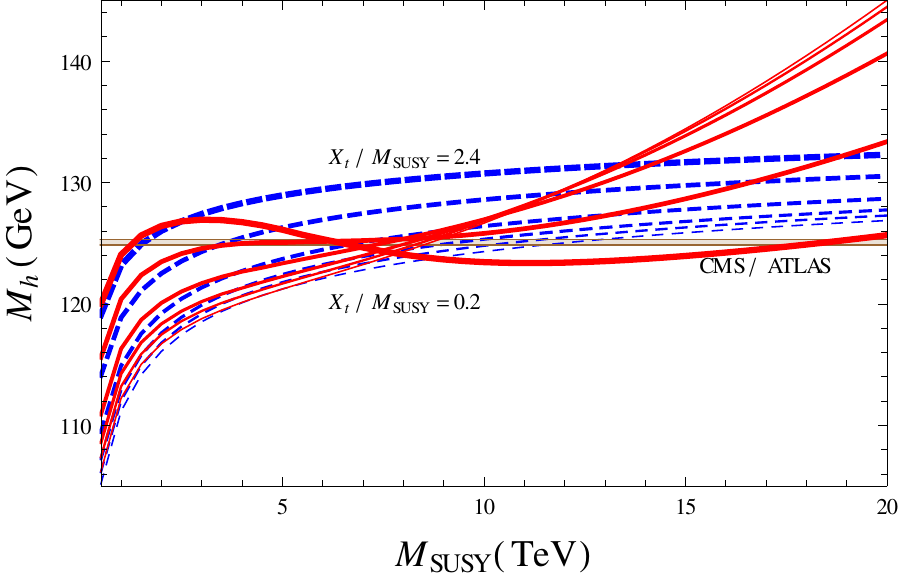} \\
\vspace*{0.5cm}
\includegraphics[width=18pc]{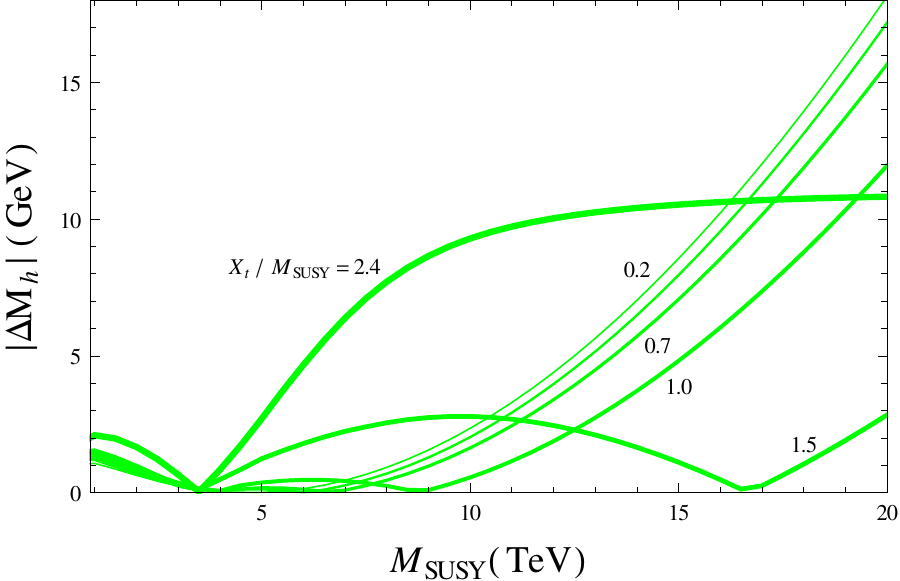}
\caption{\small{Numerical comparison of the $M_h$ predictions for a non-vanishing stop mixing in the heavy SUSY limit. The blue dashed lines are the NNLL predictions of FeynHiggs and the red solid lines represent our three-loop fixed-order predictions. The brown band is the CMS/ATLAS Higgs boson mass, $M_h^{exp}=125.09 \pm 0.24~\rm{GeV}$. Up: $M_h$ as a function of $M_{SUSY}$ for different stop mixing values, $X_t/M_{SUSY}~=~0.2,~0.5,~0.7,~1.0,~1.5~\rm{and}~2.4$. Down: Absolute numerical differences between the three-loop fixed-order predictions and the NNLO results of FeynHiggs plotted in the upper figure. }}
\label{Fig3}
\end{figure}\\
We further explore the dependence of the Higgs boson mass on the SUSY input parameters $M_{SUSY}$, $X_t$ and $tan\beta$ in the heavy SUSY limit. The figures \ref{Fig1} - \ref{Fig3} show that the predicted value of $M_h$ grows when $M_{SUSY}$ increases and reach a maximum value at the critical point $X_t/M_{SUSY}=2.4$, whose location is independent of $M_{SUSY}$. It suggests that one can find boundaries for the region of rMSSM parameters which put further constraints on $M_h$. Fig. \ref{Fig5} shows the numerical values of $X_t/M_{SUSY}$ and $M_{SUSY}$ which produce the same Higgs mass prediction (gray curves). We have considered values of $M_h$ from $115~\rm{GeV}$ to $131~\rm{GeV}$ and set $tan\beta = 10$. We observed here that there is a minimum value of $M_{SUSY}$, located at the maximal point $X_t/M_{SUSY} = 2.4$, which is compatible with some election of the Higgs boson mass. Moreover, in the case of non stop mixing ($X_t=0$) one can find the higher value of $M_{SUSY}$ compatible with a given $M_h$. These extrema values grow when we consider higher values of $M_h$. This behaviour can also be seen at the intersection of the brown band with the blue dashed lines in Fig. \ref{Fig3} for a $125~\rm{GeV}$ Higgs mass.
\begin{figure}
\centering
\includegraphics[width=18pc]{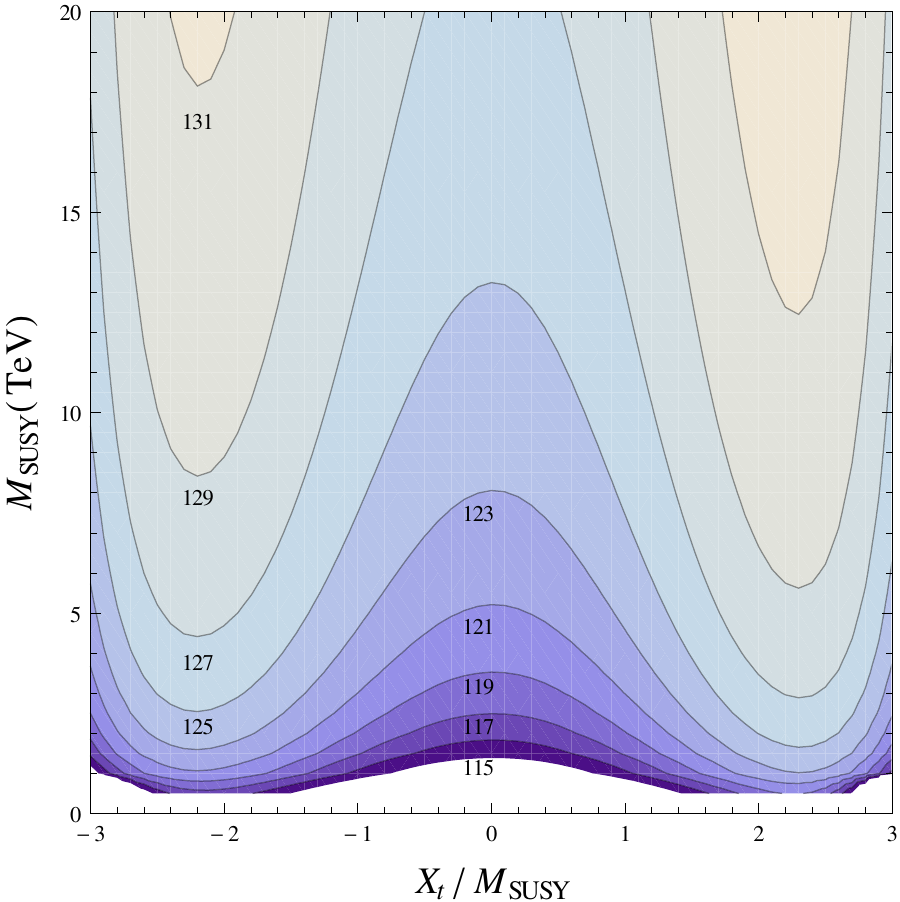}
\caption{\small{Dependence of $M_h$ on $M_{SUSY}$ and $X_t$ in the heavy SUSY limit. We have used $tan\beta = 10$. The gray lines represent the values of $M_{SUSY}$ and $X_t$ which produce the same Higgs boson mass. The predicted value of $M_h$ increases monotonically with $M_{SUSY}$. }}
\label{Fig5}
\end{figure}
If we use the combined CMS/ATLAS measured Higgs boson mass within the actual combined uncertainties, $M_h^{exp}=125.09 \pm 0.24~\rm{GeV}$, we will be able to fix upper and lower bounds on the SUSY scale $M_{SUSY}$ in the benchmark scenario considered in this work. \\ Fig. \ref{Fig6} shows the $125.09~\rm{GeV}$ contours (gray lines) as a function of $M_{SUSY}$, $tan\beta$ (Up: for values of $X_t/M_{SUSY}$ from 0 to 2.4) and $X_t/M_{SUSY}$ (Down: for values of $tan\beta$ from 4 to 30). The blue and the brown regions refer to the SUSY parameters compatible with $M_h^{exp}$. The purple lines represent the combined uncertainty for the cases enclosed inside. 
Notice that if $tan\beta \leq 10$ then $M_{SUSY}$ strongly depends on $tan\beta$, moreover the parameter region of $tan\beta \lesssim 3$ is incompatible with the LHC observations of the Higgs boson mass if one considers SUSY scales below 20 TeV. For values above 10, the dependence is marginal and the curves flatten. As a consequence, at low $tan\beta$ values, independent of the election of $X_t$, it is not possible to find upper bounds on the required SUSY scale from the CMS/ATLAS Higgs mass value. There is still the possibility to use the vacuum stability of the Higgs potential to find upper bounds on $M_{SUSY}$ for small $tan\beta$ values \cite{Allanach} but this is beyond the scope of the present work. For higher values however ($tan\beta\gtrsim 10$), due to the curves are almost constant, one can identify a lower bound for $X_t/M_{SUSY}=2.4$ and an upper bound for a vanishing stop mixing parameter ($X_t=0$). When $tan\beta=10$, which is the point considered in all the above plots of this section, we find that $M_{SUSY}$ must be at most $12.5\pm1.2~\rm{GeV}$ (see purple line in upper plot) in order to be in agreement with the CMS/ATLAS Higgs mass value. $M_{SUSY}$ can be reduced up to $9.6~\rm{GeV}$ for $tan\beta=30$ and $X_t=0$. 
\begin{figure}
\centering
\includegraphics[width=18pc]{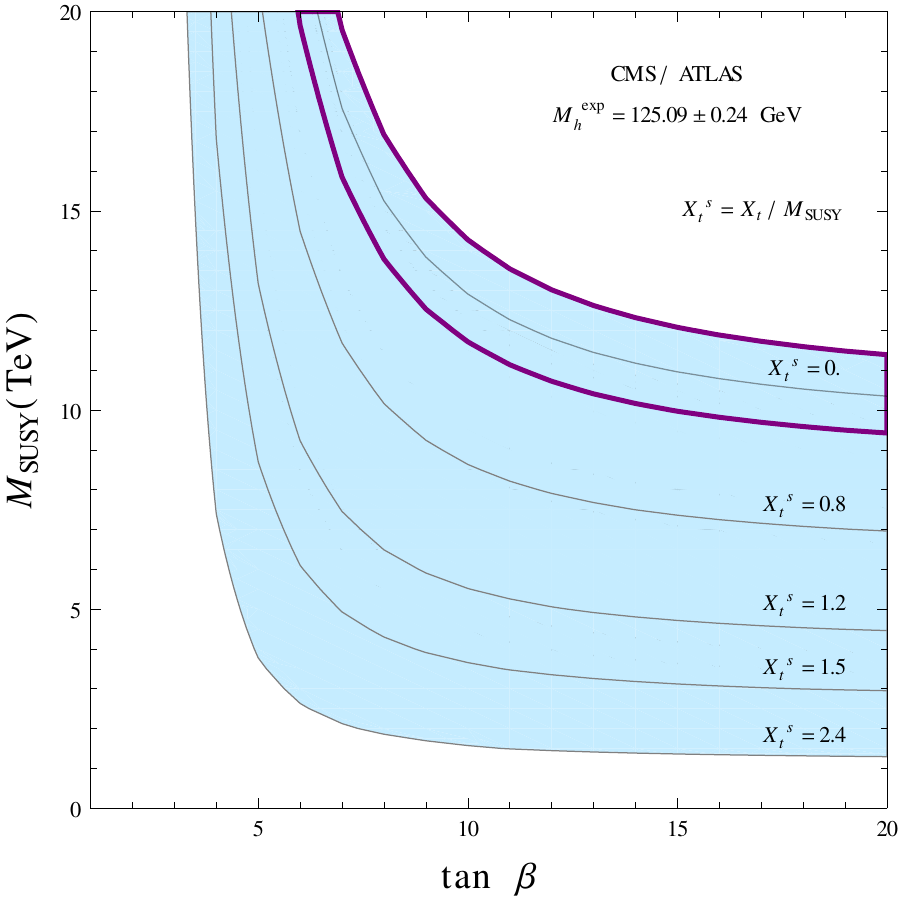} \\
\vspace*{0.5cm}
\includegraphics[width=18pc]{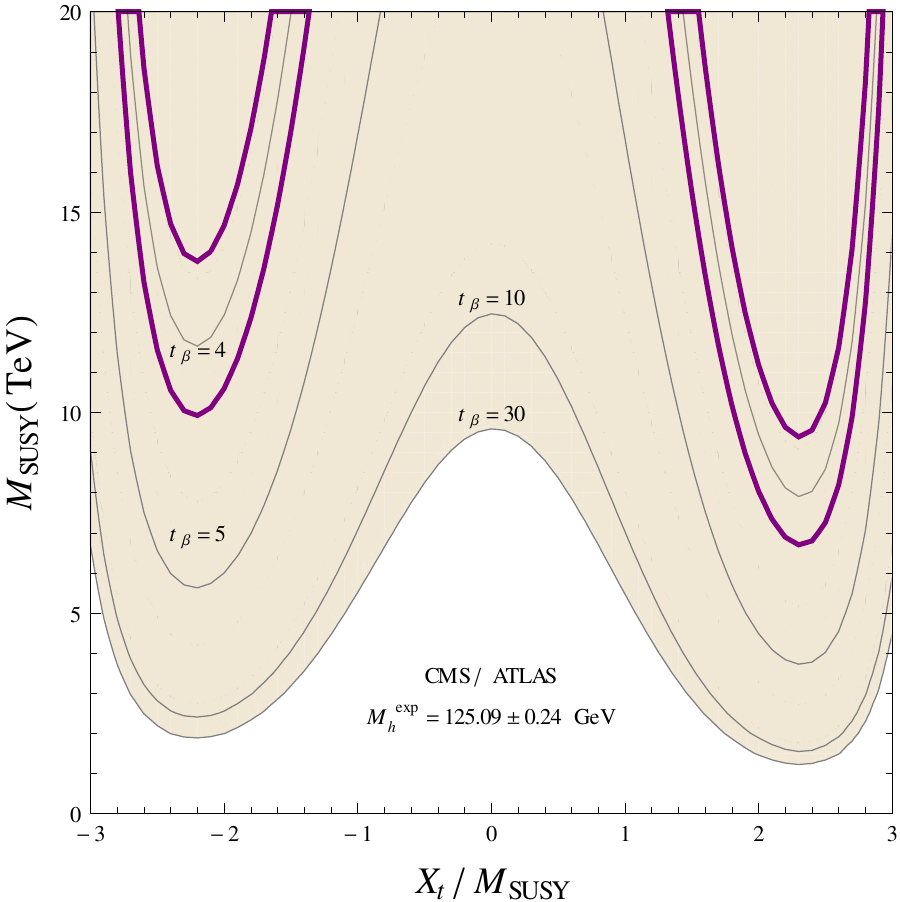}
\caption{\small{Region of rMSSM parameters in the heavy SUSY limit which is compatible with the central value and the combined uncertainty of the CMS/ATLAS Higgs boson mass, $M_h^{exp}=125.09\pm0.24~\rm{GeV}$. Up: Gray lines represent the points ($M_{SUSY}$, $tan\beta$) compatible with a $125.09~\rm{GeV}$ Higgs mass for different values of the stop mixing parameter, $X_t/M_{SUSY}=0,~0.8,~1.2,~1.5,~2.4$. The purple line represents the combined uncertainty for the case of zero stop mixing. Down: Gray lines are the 125.09 GeV contours as a function of $M_{SUSY}$ and $X_t/M_{SUSY}$ for different values of the parameter $tan\beta$, $tan\beta=4,~5,~10,~30$. The purple lines are the points compatible with the combined uncertainty for the lowest value of $tan\beta$ considered.}}
\label{Fig6}
\end{figure}
One can significantly lower the required value of $M_{SUSY}$ to $1.2~\rm{TeV}$ when $|X_t/M_{SUSY}|$ increases up to $2.4$ and for $tan\beta=30$. The region $M_{SUSY}>12.5\pm1.2~\rm{TeV}$, where the three-loop fixed-order results blow up, is excluded by the combined CMS/ATLAS Higgs mass value in the simple scenario consider here. The coming combined result for RUN 2 by ATLAS and CMS will reduce the current uncertainty and therefore the upper bound on the SUSY scale (for higher values of $tan\beta$) could be reduced even more.                      

\section{\large Conclusions}
\label{sec-4}

We have recently presented a fixed-order computation of the lightest rMSSM Higgs boson mass which extends the validity of the leading three-loop corrections to the whole parameter space of the rMSSM~\cite{Edilson}. This computation is in a very good agreement with the results of H3m~\cite{Harlander3} for low SUSY scales ($M_{SUSY} \lesssim 1.2~\rm{TeV}$). However for large $M_{SUSY}$ a numerical comparison with the available codes is missing. We have decided to filling this gap by checking our computation of $M_h$ with the three-loop results coming from the EFT hybrid approach implemented in FeynHiggs~2.14~\cite{FeynHiggs3} for the same observable. FeynHiggs includes the resummation of the large logarithms at high SUSY scales and is in a very good agreement with the other fixed-order and EFT codes. This allowed us to compare our results with a reliable three-loop $M_h$-prediction for $M_{SUSY}$ up to 20 TeV. We focused on a single SUSY scale scenario in the decoupling limit (heavy SUSY limit) where the SM is the low energy EFT. We specifically compared our $O(\alpha_t\alpha_s^2)$ and the FeynHiggs NNLL predictions of $M_h$ at the kinematical point $A_{e,\mu,\tau,u,d,c,s,b}=0$, $tan\beta=10$ and $\mu_r=M_{SUSY}$. We find a very good agreement between the two results for SUSY scales below 10~TeV in the case of vanishing stop mixing ($X_t=0$). This agreement can be improved for a different election of the parameters $M_{\tilde{g}}$ and $X_t$. The difference is estimated to be in the range $ 0.2~GeV \lesssim \Delta M_h \lesssim 1~\rm{GeV}$ for the region $M_{SUSY}\lesssim 10~\rm{TeV}$. Above $M_{SUSY}=10~\rm{TeV}$ we have observed significant differences that increase monotonically with $M_{SUSY}$. Such a behaviour is expected for high SUSY scales since the $O(\alpha_t\alpha_s^2)$ computation contain the effects of the large logarithms. Nevertheless, the region where the contributions of the large logarithms blow up is excluded by the combined CMS/ATLAS Higgs mass, $M_h^{exp}=125.09 \pm 0.24~\rm{GeV}$. We have derived an upper bound on the needed SUSY scale for the considered scenario. For values above $tan\beta=10$ the region $M_{SUSY}>12.5\pm1.2~\rm{TeV}$ is ruled out. \\

\vspace*{0.2cm}

\textbf{Acknowledgements.} This work is partially financially supported by the research grant N. 39844 of the call CONVOCATORIA 727 DE COLCIENCIAS PARA DOCTORADOS NACIONALES 2015.

\end{document}